\documentstyle[floats,aps,prl]{revtex}
\begin{document}
\input epsf
\draft
\twocolumn[\hsize\textwidth\columnwidth\hsize\csname  
@twocolumnfalse\endcsname
\bibliographystyle{}

\title{Phase shift effects on the second moment and skewness of the field
profiles obtained by the muon spin
relaxation technique}
\author{X. Wan}
\address{Physics Department, College of William and Mary, Williamsburg, 
VA 23187}
\date{\today}
\maketitle

\begin{abstract}
Abstract. Recent high transverse field muon spin relaxation (TF-$\mu $SR)
experiments performed on Bi2212 single crystals show that the phase
parameter extracted from individual histogram,
an indication of the angle between the
muon's initial polarization direction and the detector direction, is
temperature dependent. This phase shift effect, probably due to
the electronic or initial polarization instability at a time interval
comparable with the
muon spin precession period, will affect the second moment and skewness of
the field profiles. The proper data analysis procedure is discussed to
correct this phase shift effect, which is important on the
quantitative analysis of the
first order transition or melting transition in the Bi2212 mixed state.     

\end{abstract}

\vskip2pc]  
 
\narrowtext 

\section{Introduction}

In a $\mu $SR experiment \cite{Brewer} the spin-polarized muons are
assumed to
stop at random positions within the sample and precess along the local
(internal) magnetic fields $\vec{B}(\vec{r})$. The procession frequency is
$\gamma _{\mu }B$,
where $\gamma _{\mu }=135.5$ $MHz/T$ is the gyromagnetic ratio of the
muon. If the local fields which the muons feel are not unique i.e. there
is a
distribution of magnetic field $f(B)$,
consequently the
muons will precess with different frequencies, which causes a
depolarization of the muons ensemble's spins. The second moment of the   
field distribution $f(B)$ indicates how fast the muons depolarize thus is
called the relaxation rate ($\sigma$ or $\lambda$). There are several ways
to extract the
relaxation rate: we can assume the proper form of
the depolarization
function $P(t)$ to fit the $\mu $SR spectra (histogram or asymmetry) in
the
time domain to extract 
the relaxation rate parameter in the fitting procedure; or we can calculate 
the second moment of the
field distribution $f(B)$ to get the relaxation rate $\sigma$ since the
Cosine Fourier
Transformation (CFT) of $\mu $SR spectra indicates the magnetic field  
distribution inside the sample. \cite{Brewer,Lee} The second moment
calculated from the discretized field distribution data depends
slightly on the field channel width $\Delta$, there exists the following 
relations: \cite{Wan}
\begin{eqnarray}
\sigma ^{2} &=&\frac{\int b^{2}f(b)db}{\int f(b)db}-\left( \frac{\int
bf(b)db%
}{\int f(b)db}\right) ^{2}  \nonumber \\
&\approx &\frac{\sum b^{2}f(b)db}{\sum f(b)db}-\left( \frac{\sum
bf(b)db}{%
\sum f(b)db}\right) ^{2}-\frac{\Delta ^{2}}{12}  \nonumber \\
&=&\sigma _{data}^{2}-\frac{\Delta ^{2}}{12}
\end{eqnarray}
where $\sigma$ is the true relaxation rate and $\sigma_{data}$ is the
relaxation rate calculated from the discretized field profile. Simulation
shows that as long as the channel width $\Delta$ is much 
smaller than the second moment of the field profile, $\sigma_{data}$ is a
good estimate of $\sigma$. \cite{Wan} For a $\mu $SR spectra
in a 10 $\mu$s time   
window, the ideal minimum relaxation rate we can extract from data is
0.628 $\mu$s$^{-1}$. If we consider the channel width effect mentioned
above, the resolution of the relaxation rate will still be larger. \\
The local fields $f(B)$ can either be
intrinsic, as they are
for ordered magnets and spin glasses, or induced by an external field, as
for
the vortex lattice of a type II superconductor formed in the external
magnetic fields. Usually the distribution $f(B)$ is dependent
on the flux distribution associated with
a single vortex line, as well as the arrangement and dynamics of the
vortex lines. Note for single vortex, the magnetic field
has a
component normal to the vortex axis, however, the average transverse     
component vanishes due to the large amount of contributions from the
entire FLL \cite{Kogan}. For general calculation of the magnetic field
distribution, see the monograph of Greer and Kossler \cite{Greer}.
\section{Phase-shifted $\mu $SR spectra}
In most cases, there is no difficulty to obtain the true field profiles
through CFT
from low 
transverse field $\mu $SR data. However, when the
applied magnetic field is so high (several Tesla) that one precession
period is only
several times of the channel width, we need to study the effect of
time-shifted or phase-shifted $P(t)$ (due to the instabilities of
electronics or initial polarization) on the
interpretation of $f(B)$. Assume $P(t)=G_{T}(t)\cos (\omega _{0}t)$ where 
$G_{T}(t)$ is the   
non-oscillating depolarization function part and can be exponential,
stretched
exponential, Gaussian or other types. Use the notations
\begin{eqnarray}
P_{c}(\omega )&=&\int G_{T}(t)\cos (\omega t)\text{ }dt \\ 
P_{s}(\omega )&=&\int G_{T}(t)\sin (\omega t)\text{ }dt
\end{eqnarray}
We can easily derive the Fourier Transform (FT) of the time-shifted (or
phase-shifted) $P(t)$
as follows:
\begin{eqnarray}
P_{sc}(\omega ) &=&\int G_{T}(t+\delta t)\cos [\omega     
_{0}(t+\delta t)]\cos (\omega t)\text{ }dt  \nonumber \\
&=&\frac{\cos (\omega _{0}\delta t)}{2}P_{c}(\omega _{0}-\omega
)-\frac{%
\sin (\omega _{0}\delta t)}{2}P_{s}(\omega _{0}-\omega )
\end{eqnarray}
\begin{eqnarray}
P_{ss}(\omega )=-\frac{\sin (\omega _{0}\delta t)}{2}P_{c}(\omega _
{0}-\omega )-\frac{\cos (\omega _{0}\delta t)}{2}P_{s}(\omega
_{0}-\omega )
\end{eqnarray}
Note in the expression of $P_{sc}(\omega )$, the 1st term is
symmetric about $\omega _{0}$, however, there is a sign change in the 2nd
term when $\omega $ goes across $\omega _{0}$. This sign change was seen 
in our Bi2212 high transverse field $\mu $SR data \cite{Wan} at
various temperatures. \\
To construct the true field profile $P_{c}(\omega _{0}-\omega )$ from
the experimental data $P_{sc}(\omega )$ and $P_{ss}(\omega )$, we can 
do the following transformation,
\begin{eqnarray}
P_{sc}(\omega )\cos (\omega _{0}\delta t)-P_{ss}(\omega
)\sin (\omega _{0}\delta t)=\frac{1}{2}P_{c}(\omega _{0}-\omega )
\end{eqnarray}
The result of this phase correction on one of the field profiles in
Bi2212 is shown
in figure 1 where we can see the proper field profile is obtained. \\
\begin{figure}[h]
\centerline{\epsfxsize=2.375in \epsfbox{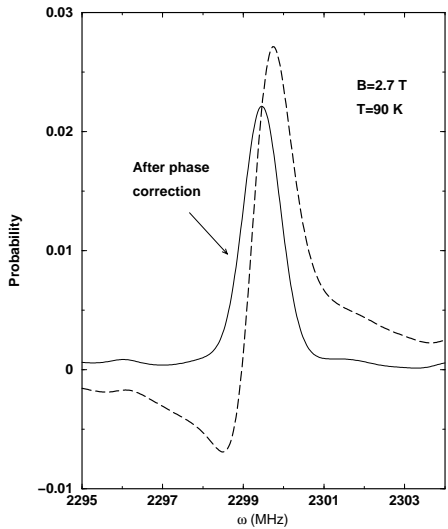}}
\vskip .25in  
\caption{The field profile on the {\it ab} basal plane of Bi2212 obtained
before (the dashed line) and after (the solid line) the phase correction 
as discussed in the text.}
\end{figure}
The measurement of the internal magnetic field
distribution in the vortex of type II HTSC allows an investigation of
the details of the static or dynamic flux structure \cite{Lee}, such as
the effects of random pinning \cite{Brandt} and the
dimensionality/melting of the vortex \cite{Brandt,Bernhard}. To
quantify and characterize the shape of the field distribution (field
profile), skewness 
factor $\alpha $, the
variation of which reflects the underlying changes of the vortex
structure (the one-to-one mapping still needs to be studied further), 
is defined from the 3rd and 2nd moments of the field line shape as
follows:
\begin{eqnarray}
\alpha  &=&\frac{\langle (B-\overline{B})^{3}\rangle
^{\frac{1}{3}}}{\langle
(B-\overline{B})^{2}\rangle ^{\frac{1}{2}}}   \\
&=&\frac{(M_{3}-3M_{1}M_{2}+2M_{1}^{3})^{\frac{1}{3}}}{(M_{2}-M_{1}^{2})^{%
\frac{1}{2}}}
\end{eqnarray}
where $M_{n}=\int B^{n}f(B)dB.$ and is the nth moment of $f(B)$. \\
Normally we obtain the field profile either from  
asymmetry plot or from individual histogram then calculate the second
moment and skewness of the field profile to probe the possible phase
transitions of the vortex matters. In the analysis of high transverse
field $\mu $SR data on
Bi2212, we
often find that the phase parameter in the depolarization function changes
with
temperature and thus needs to be corrected precisely. 
\section{Calculation results}
To evaluate the effects of the phase shift $\phi $ on the
second moment and skewness calculations of the field profile, we generate
a
depolarization function (asymmetry plot) $P(t)=e^{-\frac{\sigma
^{2}t^{2}}{2}%
}\cos (\omega t+\phi )$ where $\sigma =2$ $\mu s^{-1}$, $\omega =20$
$Mrad/s$ ($\omega $ can be assumed much higher, the final results are
similar).
These parameters are chosen so that there are no evident visual changes in
the
field profiles when the phase shift $\phi $ varies from 0 to 0.5 ($\sim
$30$%
^{0}$). The calculated $\phi $ dependent second moment (ideally the second
moment  
should be $2$ $\mu s^{-1}$) and skewness plot is shown in the figure 2. \\
\begin{figure}[h]
\centerline{\epsfxsize=2.375in \epsfbox{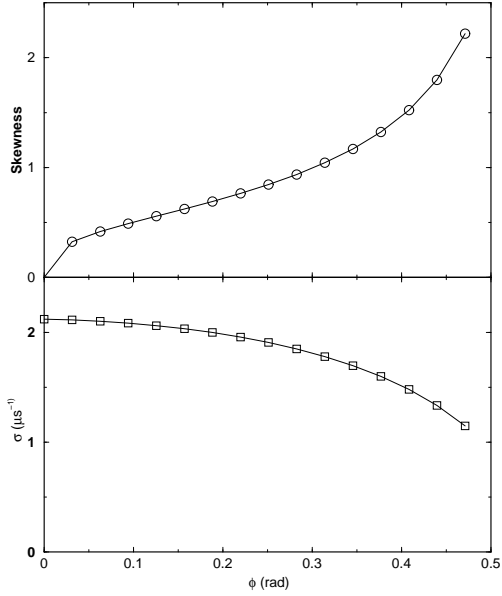}}
\vskip .25in
\caption{The phase shift effects on the second moment and skewness of
the field profile, detailed procedure is discussed in the text.}
\end{figure}
We can see clearly that the second moment and skewness are very sensitive
on
the phase shift $\phi $. A change of initial phase from 0 to 0.5 ($\sim
$30$%
^{0}$) causes about 50 percent change on the relaxation rate and an
increase of skewness from 0 to 2. This result
requests us to extract the phase
parameter accurately after fitting the $\mu $SR asymmetry plot. To correct
this phase-shift effect, we can follow the modified CFT procedure as
discussed in the previous section to obtain the
true field profiles, which is also the case when we analyze the individual
histogram. \\
We now study the asymmetry $A(t)$ which is used to obtain the field
profile. We know $A(t)$ is usually composed by two histograms coming
from two
oppositely ($\pi $ out of phase) located positron detectors. However, in
reality,
especially for high precession frequency data, the two histograms are
usually found not exactly $\pi $ out of phase. To consider this phase
deviation $\phi $ from $\pi $, we can do the following analytical
analysis.
As usual, we assume the phase-shifted histograms are (we assume Gaussian
depolarization here, we can also use other depolarzation function instead)
\begin{eqnarray}
N_{U}(t)&=&N_{0}e^{-\frac{t}{\tau }}[1+A_{0}e^{-\frac{\sigma
^{2}t^{2}}{2}%
}\cos (\omega t)]  \\
N_{D}(t)&=&N_{0}e^{-\frac{t}{\tau }}[1-A_{0}e^{-\frac{\sigma
^{2}t^{2}}{2}%
}\cos (\omega t+\phi )]
\end{eqnarray}
The asymmetry $A(t)$ can be expressed as \\
\begin{eqnarray}
A(t) &=&\frac{N_{U}-N_{D}}{N_{U}+N_{D}}  \nonumber \\
&=&\frac{A_{0}e^{-\frac{\sigma ^{2}t^{2}}{2}}\cos (\omega   
t)+A_{0}e^{-\frac{%
\sigma ^{2}t^{2}}{2}}\cos (\omega t+\phi )}{2+A_{0}e^{-\frac{\sigma
^{2}t^{2}%
}{2}}\cos (\omega t)-A_{0}e^{-\frac{\sigma ^{2}t^{2}}{2}}\cos (\omega
t+\phi
)}  \nonumber \\
&=&\frac{A_{0}e^{-\frac{\sigma ^{2}t^{2}}{2}}\cos (\omega t+\frac{\phi
}{2}%
)\cos (\frac{\phi }{2})}{1+A_{0}e^{-\frac{\sigma ^{2}t^{2}}{2}}\sin
(\omega
t+\frac{\phi }{2})\sin (\frac{\phi }{2})}  \nonumber \\
&\simeq &A_{0}e^{-\frac{\sigma ^{2}t^{2}}{2}}\cos (\omega t+\frac{\phi
}{2}%
)\cos (\frac{\phi }{2})-\frac{A_{0}^{2}}{4}e^{-\sigma ^{2}t^{2}}\sin 
(2\omega t+\phi )\sin \phi
\end{eqnarray}  

From equation 11, we can see that, to the first order, the relaxation
function is composed of two depolarized harmonics with frequencies
$\omega $
and $2\omega $. The asymmetry amplitude for the $\omega $ harmonic is $%
A_{0}\cos (\frac{\phi }{2})$, which is reduced by a factor of $\cos   
(\frac{%
\phi }{2})$ from the original amplitude $A_{0}$, there is a phase shift $%
\frac{\phi }{2}$, which causes the changes on the second moment and
skewness
of the field profile as we have discussed before; since $A_{0}$ is usually
less than 0.4, the asymmetry amplitude for the $2\omega $ harmonic
$\frac{%
A_{0}^{2}}{4}\sin \phi $ is much less than $A_{0}\cos (\frac{\phi }{2})$,
the amplitude of the $\omega $ harmonic. All these conclusions have been
verified through further simulation.
\section{Conclusion}
In order to obtain the proper field profiles through the Cosine Fourier
Transform (CFT) on the $\mu $SR data and characterize them quantitatively 
and correctly, we need to extract the initial phase parameter
precisely from
individual histogram or asymmetry plot and correct the possible phase
shift
effect
through a proposed data analysis procedure in order to obtain the correct
second moment
(or
relaxation rate) and skewness of the field profile. 
\bibliography{paper6}
\end{document}